# AI-Assisted Data Extraction for Systematic Reviews in Education

**Noah L. Schroeder[1]* , Chris Davis Jaldi[2] , Shan Zhang[3] & Kweku Yamoah[1]**

[1] - Department of Computer & Information Science & Engineering, University of Florida, Gainesville, Florida.
[2] - Department of Computer Science & Engineering, Wright State University, Dayton, Ohio.
[3] - School of Teaching and Learning, University of Florida, Gainesville, Florida.

Correspondence regarding this manuscript should be addressed to: Noah L. Schroeder - schroedern@ufl.edu

**Abstract**

Systematic reviews are time-consuming endeavors that require knowledgeable human reviewers to screen studies for relevance and extract data following a specific coding scheme before any analysis or synthesis can occur. Large language models (LLMs) hold promise for substantially accelerating this process and reducing reviewer workload, yet their application within the context of systematic reviews in the field of education remains underexplored. We address this issue in two ways: through empirical studies and the iterative development of an open-source software tool. First, we conducted two empirical studies examining the efficacy of using LLMs for data extraction using data from a published review on pedagogical agents. We extracted a variety of data types from 112 studies and compared the results to data extracted by human coding. Results indicate that LLMs struggled with extracting data accurately and therefore are not ready to be used as primary data extraction tools without explicit human validation of the data extracted. These findings highlight the dire need for a human-in-the-loop (HIL) approach to AI-assisted data extraction. We then propose a HIL workflow and introduce and describe the development of a free, web-based, open-source tool designed to support user-friendly, human-validated data extraction with LLMs.

## Introduction

Large language models (LLMs) are revolutionizing the way we think about research. In educational contexts, researchers have begun investigating if LLMs can be used to generate tailored learning content (Jaldi et al., 2025; Leiker et al., 2023) or if they can be used as temporary teachers (Al Ghazali et al., 2024). As research synthesists, we recognize that ignoring the advancements in LLMs could be to our own detriment. After all, systematic review methods can be intensely time-consuming, requiring human effort and expertise throughout the entire process (Chernikova et al., 2024; Wang & Luo, 2024). LLMs have been used as second screeners during abstract screening (Cao et al., 2024; Huotala et al., 2024; Issaiy et al., 2024), for data extraction (Gartlehner et al., 2024; Konet et al., 2024; Schmidt et al., 2024), and throughout the systematic review process more generally (Sami et al., 2024; Scherbakov et al., 2024).

In this paper, we focus on the data extraction step of the systematic review process. This step of systematic review and meta-analysis workflows is critical because it is the foundation from which results are derived; there is little room for error. Traditionally, research synthesists have used two independent coders to extract data from, at a minimum, a subset of studies. By doing so, researchers are able to report inter-rater agreement or inter-rater reliability statistics, such as Cohen’s kappa. This process helps ensure not only transparency in the coding processes but also the reliability of the coding process, helping build trust in the results. Unfortunately, this process is inherently time-consuming and, therefore, expensive. Our work investigates to what extent LLMs can accelerate this process.

## LLMs for Research Synthesis Data Extraction

Reviewing relevant research in the area shows very promising, although arguably preliminary, results. Researchers have found promising results for extracting data in medical contexts (Cao, Arora, et al., 2025; Gartlehner et al., 2024; Kartchner et al., 2023), with some studies showing over 90% accuracy of the data extracted (Gartlehner et al., 2024). Yet, such results are not universal. Researchers have found more nuanced findings, with high accuracy for simple citation details, but only 10%-16% accuracy when extracting complex or subjective data types (Stewart-Evans et al., 2025). The majority of studies in the field have occurred in medical and medically-adjacent fields, but LLM-based data extraction has been found to be less accurate in the social sciences (like education) than in clinical studies (Schmidt et al., 2024).

A recent large-scale study provides evidence that LLM-based data extraction in the social sciences may not be very accurate. A comprehensive study in the social sciences was conducted by Jansen et al. (2025), who used eight LLMs to extract 186 variables from over 2,000 studies in psychology. They found that 20% of all variables did not reach an acceptable level of accuracy, and they found mixed accuracy (meaning acceptable and unacceptable, depending on the metric) for 46% of the variables.

Some relevant research does exist specifically in the area of education, and a recent study showed promising results. Wang and Luo (2026) evaluated *MetaMate*, an AI-assisted data extraction software platform. They extracted 20 variables from 32 studies and found precision exceeded 80%, recall exceeded 90%, and F1 scores ranged from 88%-96%. While extracting 20 variables from 32 studies from one meta-analysis is a relatively constrained sample in this context, it demonstrates the potential of LLM-based data extraction in educational contexts. However, further replications and extensions are necessary.

To summarize, research to date shows that LLMs can sometimes extract data accurately, but the conditions under which they have done so are nuanced. Generally speaking, however, extensions of the existing research are needed with a larger corpus of studies, particularly in educational fields.

### Limitations of Existing Research Around LLMs for Data Extraction

Researchers have moved quickly in the field exploring how LLMs can extract data for systematic reviews. However, based on the extant body of research so far, there are three sets of limitations we must consider.

**LLM-specific limitations.** Cao, Arora, et al. (2025)'s results are promising, but their previous work on developing prompt templates for study screening (Cao, Sang, et al., 2025) highlights known problems with LLMs: results from one model may not replicate between models and model families, and LLMs can be sensitive to the prompt (Anagnostidis & Bulian, 2024; Errica et al., 2024) and parameter settings such as seed value, temperature, top-k sampling, etc. (Loya et al., 2023). Moreover, prompts do not transition well between domains (Schmidt et al., 2024), and LLMs perform differently even when shown the same prompt with the same datasets (Cao, Sang, et al., 2025). In addition, the risk of hallucination (Tonmoy et al., 2024) is very real. Even if hallucinations were a solved problem, some studies have found the data returned by the LLM was noisy (Kartchner et al., 2023), and one must remember that each model has different strengths and weaknesses (Kapoor & Arora, 2024; Matarazzo & Torlone, 2025).

Taken together, all these factors highlight the extensive amount of evidence that should be available in order to make generalizable claims about LLM-generated results in the research synthesis context.

**Statistical limitations.** When interpreting studies examining the efficacy of LLMs for data extraction, accuracy alone may not fully highlight the strengths and weaknesses of an algorithm's performance. When evaluating classification tasks such as this, precision, recall, and F1 score can be more informative. *Precision* reflects the proportion of cases identified by the algorithm as accurate that are indeed accurate, which indicates the reliability of its positive classifications. Meanwhile, *recall* tells us the proportion of truly accurate cases that the algorithm successfully identified, which indicates its ability to capture all relevant instances. The *F1 score* is a proportion calculated as two times the precision multiplied by the recall, divided by the sum of the precision and recall (Gartlehner et al., 2025). Together, accuracy, precision, recall, and F1 can provide a more in-depth understanding as to the algorithm's (in this case, LLMs) performance, where it did well, and where it did not. In data extraction tasks for research synthesis, the results for accuracy, precision, and recall have been varied and illuminating. For example, previous research has reported patterns of high precision (90%) but low recall (25%) (Stewart-Evans et al., 2025).

**Methodological limitations.** A fair proportion of the existing work in the area is admittedly preliminary, as small samples inherently limit the interpretation and generalizability of the findings. For example, existing work on extracting data with LLMs has used only a small number of PDFs (Gartlehner et al., 2024; Konet et al., 2024), extracted data for relatively few variables (Gartlehner et al., 2024; Kartchner et al., 2023), and explored a similar group of LLMs (generally, ChatGPT or Claude). As noted, the vast majority of the work in the field has occurred in medically-related fields. This is a very important caveat because education and other social sciences often have more complex conceptual groundings than some other fields (Schmidt et al., 2024), and conceptual complexity may make it more difficult for LLMs to differentiate between things a human reader could easily differentiate.

## The Need for Empirical Replications in Education

Based on the prior work, it is clear that LLMs offer the *potential* to facilitate the extraction of data for systematic reviews and meta-analyses in education. However, it is also clear that replications are needed in different fields if generalizable claims are to be made. We conducted a pilot study and a main study to examine the performance of LLMs for data extraction in education. Through both, we investigated the following research questions:

1. How accurately can an LLM extract explicitly stated data?
2. How accurately can an LLM extract and categorize data based on predefined categories?
3. What is the overall accuracy of an LLM's data extraction compared to a human coder?

# Pilot Study

## Overview of Pilot Study

Our initial exploration of using LLMs for data extraction took place in the Spring of 2024. At that point, LLMs were still a relatively new technology and were just beginning to become more widely adopted. We used what were three leading LLM services at the time: Claude 2.1 [1], ChatPDF [2], and GPT-4 [3]. All were used via their web-based chat interface. We extracted data from four journal articles and four conference proceedings related to the use of virtual characters in K-12 education. We asked each LLM to extract 24 variables, including data that are generally *explicitly stated* in published papers (e.g., duration of study, learning topic, etc.) and *derived data* based on our own categorization scheme (e.g., agent role, agent type, study media based on our categorization scheme, etc.). We chose to include both types of data because both are typically extracted during systematic reviews and meta-analyses. In total, we had nine explicit variables and 15 derived variables extracted from each study. Supplementary Materials 1 shows the prompt used to extract each variable. Notably, we did not try to perfect each prompt. Rather, our goal was to write prompts in a way that an average researcher approaching an LLM may write them.

In our pilot study, we calculated simple inter-rater agreement between the human-extracted data for each study and the LLM-extracted data for each study, as the goal was to determine if there was a rationale for a larger systematic study rather than generalizable claims. Supporting data are available on OSF: https://osf.io/nraxf/?view_only=879654690be249cc9237525ed7f86ec2.

## Pilot Study Results

### How accurately can an LLM extract explicitly stated data?

We asked each LLM to extract nine explicit variables from each of the eight data sources. Compared to the human coder, we found that Claude was the most accurate LLM, showing 93.06% agreement with the human coder. GPT-4 was also quite accurate (91.67% agreement), and ChatPDF was reasonably accurate (75% agreement).

### How accurately can an LLM extract and categorize data based on predefined categories?

We also asked each LLM to extract 15 derived variables from each of the eight data sources. We found that Claude was again the most accurate (80.00% agreement). GPT-4 and ChatPDF performed notably worse (64.17% agreement and 60.00% agreement, respectively).

1 https://claude.ai/

2 https://www.chatpdf.com/

3 https://openai.com/

### What is the overall accuracy of an LLM's data extraction compared to a human coder?

Finally, we examined the overall accuracy of the LLM's data extraction compared to that of a human coder. Claude was the best performing (84.90% agreement), followed by GPT-4 (74.48% agreement) and ChatPDF (65.63% agreement).

## Implications of Pilot Study Results

We found that the LLMs' ability to extract data was greatly dependent on the type of data being extracted and the prompt used. While the accuracy of our data extraction was not as high as that seen in other studies (Gartlehner et al., 2024), we still found promising results from leading LLMs.

# Main Study

In late 2024, we re-initiated our investigation with a more thorough main study using state-of-the-art LLMs at the time. We investigated the same research questions as in the pilot study.

## Methods

### LLMs Used

The LLMs used in the main study were Google's Gemini 1.5 Flash, Gemini 1.5 Pro, and Mistral Large 2. Notably, all three of these LLMs were available via API free of charge at the time. The choice of free models rather than paid models was intentional, as we were curious about the performance of these models for unfunded projects and those in low-resource environments. At the time the study was conducted, Gemini 1.5 Pro and Mistral Large 2 were the leading models by those respective providers.

### Data Set

The dataset consisted of 112 studies included in a published scoping review (Zhang et al., 2024).

### Data Extraction

For this main study, we used an LLM to help generate Python code for using the LLM APIs to extract the data from the studies in our sample. The Python files are available on OSF. The prompts are available as Supplementary Materials 1.

### Data Analysis

The data analysis began by using the same methods as in the pilot study. In addition, we also classified responses as an "Accurate Match", but not identical to human coding, and reported that data separately. We report the simple agreement statistic as well as Cohen's Kappa. In addition, we report the precision, recall, and F1 score for each LLM. For Kappa, precision, recall, and F1 score, 1 is the highest (max) score possible. Kappa ranges from -1 to 1, while precision, recall, and F1 score range from 0-1.

### Data Availability

All data, scripts, and analysis files are available on OSF: https://osf.io/nraxf/?view_only=879654690be249cc9237525ed7f86ec2.

## Results

We found that the Gemini 1.5 Pro and Gemini 1.5 Flash models were able to analyze all 112 studies in our sample. Moreover, the results were presented relatively consistently (see raw data extraction in the OSF repository). Mistral only returned results for 103 of the studies, and of those, three were incomplete, and one had somewhat irrelevant results. Consequently, our results for Gemini 1.5 models are for 112 studies. For Mistral Large 2, we counted the studies that were not returned or analyzed as *not matching*, meaning we still calculated the results as if we had received results for the full sample. As a consequence, the Mistral Large 2 scores are lower because there were a number of studies not reported, meaning for each study not reported, there are 24 variables not reported, or in this context, not matching human coding.

In an effort to provide more thorough statistical reporting than the pilot study, our analyses include reporting the *precision* (the proportion of cases identified by the algorithm as accurate that are indeed accurate) and *recall* (the proportion of truly accurate cases identified) and *F1 score* (proportion calculated as two times the precision multiplied by the recall, divided by the sum of the precision and recall).

### How accurately can an LLM extract explicitly stated data?

We asked each LLM to extract nine explicit variables from all 112 data sources. We further divided the comparison metrics into two categories: Exact Match and Accurate Match. The results are presented in Figure 1.

Our analysis of *exact matches* for explicitly stated data showed the following: Google Gemini 1.5 Pro was the most accurate LLM (83.23% agreement, k = 0.40), followed by Gemini 1.5 Flash (81.55% agreement, k = 0.39) and Mistral Large 2 (71.43% agreement, k = 0.30). While precision was quite high, ranging from 0.89 (Gemini 1.5 Pro) through 0.95 (Gemini 1.5 Flash, Mistral Large 2), recall ranged more widely from 0.67 (Mistral Large 2) to 0.88 (Gemini 1.5 Pro). The confusion matrices show that, generally speaking, models were more likely to extract inaccurate data rather than claim the information was not reported when it actually was.

The *accurate match* metrics showed that Google Gemini 1.5 Pro had the highest accuracy (85.02% agreement, k = 0.41), followed by Google Gemini 1.5 Flash (82.64% agreement, k = 0.39) and Mistral Large 2 (74.01% agreement, k = 0.32). Again, we see that Kappa did not reach a level typically acceptable in research synthesis contexts, but precision remained high (ranging from 0.90 - 0.95). Recall was similar to the results from the accurate match metrics (ranging from 0.70 - 0.90). The confusion matrices show that, generally speaking, models were more likely to extract inaccurate data rather than claim the information was not reported when it actually was.

**Exact Match to Human Coding**

**Gemini 1.5 Flash**

| Human Label | LLM Prediction: Incorrect | LLM Prediction: Correct |
| --- | --- | --- |
| Not Reported | 3.17% | 23.31% |
| Data Extracted | 15.18% | 58.33% |

| | |
| --- | --- |
| Agreement: | 81.55% |
| Kappa: | 0.39 |
| Precision: | 0.95 |
| Recall: | 0.79 |
| F1 Score: | 0.86 |

**Gemini 1.5 Pro**

| Human Label | LLM Prediction: Incorrect | LLM Prediction: Correct |
| --- | --- | --- |
| Not Reported | 7.84% | 18.75% |
| Data Extracted | 8.83% | 64.58% |

| | |
| --- | --- |
| Agreement: | 83.23% |
| Kappa: | 0.40 |
| Precision: | 0.89 |
| Recall: | 0.88 |
| F1 Score: | 0.89 |

**Mistral Large 2**

| Human Label | LLM Prediction: Incorrect | LLM Prediction: Correct |
| --- | --- | --- |
| Not Reported | 2.98% | 19.54% |
| Data Extracted | 25.79% | 51.69% |

| | |
| --- | --- |
| Agreement: | 71.43% |
| Kappa: | 0.30 |
| Precision: | 0.95 |
| Recall: | 0.67 |
| F1 Score: | 0.78 |

**Accurate Match**

**Gemini 1.5 Flash**

| Human Label | LLM Prediction: Incorrect | LLM Prediction: Correct |
| --- | --- | --- |
| Not Reported | 3.17% | 23.31% |
| Data Extracted | 14.09% | 59.42% |

| | |
| --- | --- |
| Agreement: | 82.64% |
| Kappa: | 0.39 |
| Precision: | 0.95 |
| Recall: | 0.81 |
| F1 Score: | 0.87 |

**Gemini 1.5 Pro**

| Human Label | LLM Prediction: Incorrect | LLM Prediction: Correct |
| --- | --- | --- |
| Not Reported | 7.74% | 18.75% |
| Data Extracted | 7.14% | 66.37% |

| | |
| --- | --- |
| Agreement: | 85.02% |
| Kappa: | 0.41 |
| Precision: | 0.90 |
| Recall: | 0.90 |
| F1 Score: | 0.90 |

**Mistral Large 2**

| Human Label | LLM Prediction: Incorrect | LLM Prediction: Correct |
| --- | --- | --- |
| Not Reported | 2.68% | 19.54% |
| Data Extracted | 23.51% | 54.27% |

| | |
| --- | --- |
| Agreement: | 74.01% |
| Kappa: | 0.32 |
| Precision: | 0.95 |
| Recall: | 0.70 |
| F1 Score: | 0.81 |

Figure 1. LLM results for data extracted that were explicitly stated in the papers.

### How accurately can an LLM extract and categorize data based on predefined categories?

We also asked each LLM to extract 15 derived variables from all 112 data sources and analyzed the responses using the same two accuracy metrics. The results are presented in Figure 2.

When evaluating the models' abilities to extract data that *exactly matches* the human coder's data extraction, Google Gemini 1.5 Pro was again the most accurate, with 65.48% agreement ($k = 0.24$), followed by Google Gemini 1.5 Flash with 64.88% agreement ($k = 0.23$) and Mistral Large 2 (56.79% agreement, $k = 0.12$). Again, we observed strong precision scores (ranging from 0.90 - 0.92), with lower recall (ranging from 0.58 - 0.70). The confusion matrices show that, generally speaking, models were more likely to extract inaccurate data rather than claim the information was not reported when it actually was.

For the *accurate match* analysis, we found that Google Gemini 1.5 Pro again performed best (70.24% agreement, $k = 0.29$), followed by Google Gemini 1.5 Flash (67.74% agreement, $k = 0.26$) and Mistral Large 2 (61.90% agreement, $k = 0.19$). Similar to the previous result, precision was high (ranging from 0.91 - 0.93) while recall was lower (ranging from 0.64 - 0.75). The confusion matrices show that, generally speaking, models were more likely to extract inaccurate data rather than claim the information was not reported when it actually was.

**Exact Match to Human Coding**

| Human Label | Gemini 1.5 Flash: LLM Prediction Incorrect | Gemini 1.5 Flash: LLM Prediction Correct | Gemini 1.5 Pro: LLM Prediction Incorrect | Gemini 1.5 Pro: LLM Prediction Correct | Mistral Large 2: LLM Prediction Incorrect | Mistral Large 2: LLM Prediction Correct |
|---|---|---|---|---|---|---|
| Not Reported | 5.60% | 2.62% | 6.79% | 1.55% | 4.58% | 2.56% |
| Data Extracted | 29.40% | 62.38% | 27.74% | 63.93% | 38.69% | 54.17% |

| | Gemini 1.5 Flash | Gemini 1.5 Pro | Mistral Large 2 |
|---|---|---|---|
| Agreement: | 64.88% | 65.48% | 56.79% |
| Kappa: | 0.23 | 0.24 | 0.12 |
| Precision: | 0.92 | 0.90 | 0.92 |
| Recall: | 0.68 | 0.70 | 0.58 |
| F1 Score: | 0.78 | 0.79 | 0.71 |

**Accurate Match**

| Human Label | Gemini 1.5 Flash: LLM Prediction Incorrect | Gemini 1.5 Flash: LLM Prediction Correct | Gemini 1.5 Pro: LLM Prediction Incorrect | Gemini 1.5 Pro: LLM Prediction Correct | Mistral Large 2: LLM Prediction Incorrect | Mistral Large 2: LLM Prediction Correct |
|---|---|---|---|---|---|---|
| Not Reported | 5.60% | 2.62% | 6.79% | 1.55% | 4.58% | 2.56% |
| Data Extracted | 26.55% | 65.24% | 22.98% | 68.69% | 33.57% | 59.29% |

| | Gemini 1.5 Flash | Gemini 1.5 Pro | Mistral Large 2 |
|---|---|---|---|
| Agreement: | 67.74% | 70.24% | 61.90% |
| Kappa: | 0.26 | 0.29 | 0.19 |
| Precision: | 0.92 | 0.91 | 0.93 |
| Recall: | 0.71 | 0.75 | 0.64 |
| F1 Score: | 0.80 | 0.82 | 0.76 |

*Figure 2.* LLM results for data extracted and categorized by LLMs from descriptions in the papers.

**What is the overall accuracy of an LLM's data extraction compared to a human coder?**

Finally, we examined the overall accuracy of each LLM's data extraction. The results are presented in Figure 3.

For the *exact match* analysis, Google Gemini 1.5 Pro led the group (72.14% agreement, k = 0.31), Google Gemini 1.5 Flash placed second (71.13% agreement, k = 0.30), and Mistral Large 2 was third (62.28% agreement, k = 0.20). Precision was high (ranging from 0.90 - 0.93) while recall was substantially lower (ranging from 0.61 - 0.76). The confusion matrices show that, generally speaking, models were more likely to extract inaccurate data rather than claim the information was not reported when it actually was.

When examining the *accurate match* analysis, Google Gemini 1.5 Pro again led the LLMs (75.78% agreement, k = 0.34), followed by Google Gemini 1.5 Flash (73.33% agreement, k = 0.32) and Mistral Large 2 (66.44% agreement, k = 0.25). Consistent with previous analyses, precision was high (ranging from 0.90 - 0.94) and recall was lower (ranging from 0.66 - 0.80). The confusion matrices show that, generally speaking, models were more likely to extract inaccurate data rather than claim the information was not reported when it actually was.

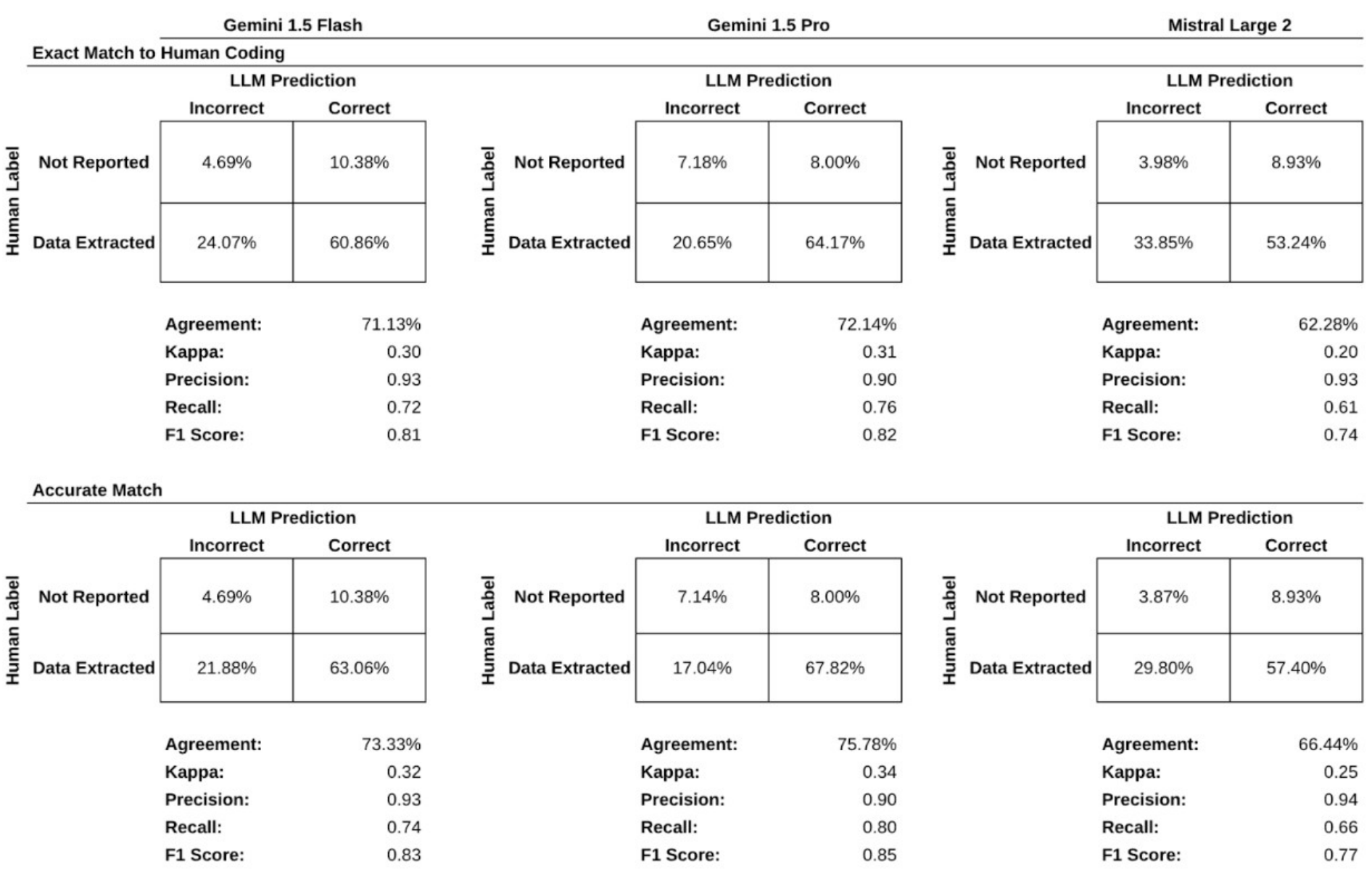

Figure 3. LLM results for all data extracted in the main study.

## Main Study Discussion

The LLMs tested here were quite accurate when extracting explicitly stated data and variables from studies, but they were substantially less effective at categorizing data based on predefined categories. Rather than focusing on one specific metric, we examined a range of metrics. Precision alone suggests LLM-data extraction is quite precise, but examining Cohen's Kappa with the human rater showed that the LLMs lack consistency with human coders. The confusion matrices showed the LLMs had often extracted inaccurate information at least 7% of the time (Gemini 1.5 Pro, "accurate match", explicitly stated data) and up to 39% of the time (Mistral Large 2, "exact match", derived data). Moreover, LLMs stated variables were "not reported" 3% (Mistral Large 2, "exact match", explicitly stated data) - 8% of the time (Gemini 1.5 Pro, "exact match" and "accurate match", explicitly stated data) even though the human extracted this data.

Our purpose was not to say one model is more effective than others, but rather to explore if LLMs can be *accurate enough* to play a role in the research synthesis process. Our results show that in a research synthesis context in education, LLMs should not be relied on for fully automated data extraction tasks without human oversight and validation. The lower accuracy we found compared to studies in medical contexts (Gartlehner et al., 2024) could be due to a series of reasons, such as LLM sensitivity to the prompt used and other settings of the LLM (Anagnostidis & Bulian, 2024; Errica et al., 2024; Loya et al., 2023). It is highly plausible that if one took enough time fine-tuning each individual question to include in the prompt, a higher accuracy could be achieved. However, this could be quite time-consuming, which would defeat the purpose of using an LLM to aid in data extraction.

We argue that the opaque probabilistic decisions of transformer-based LLMs, being driven by their training data and the training process used, means that the accuracy of any specific LLM for data extraction is irrelevant in a context as critical as research synthesis. Even if accuracy had reached 90% or higher, we would not advise researchers to adopt data extracted by LLMs without human validation. However, this realization that LLMs should not be trusted alone led our team to envision and build a human-in-the-loop (HIL) solution that allows one to use LLMs to extract data, but a human must validate the extracted data *before* it is recorded. In our informal use, we have found that this process can save time compared to only human-based data extraction.

## Human-In-The-Loop and AI-In-The-Loop Approaches to Scientific Workflows

With the availability of AI systems increasing, we are seeing increased discussion around hybrid intelligence and the collaboration of humans and AI (Cukurova, 2025; Dellermann et al., 2019). Specifically, we can conceptualize human and AI collaboration as humans doing the bulk of the work with AI supplements, or AI doing the bulk of the work with human validation, termed AI-in-the-loop and HIL, respectively (Dellermann et al., 2019). A HIL approach, where the AI does the bulk of the work on the first pass and then a human validates the information extracted, is the most fitting for research synthesis data extraction tasks. In this section, we demonstrate both our pilot work and the current version of our HIL, AI-assisted data extraction software. All software is open-source and free to use, and links to the GitHub repositories are provided in each section. Anecdotal feedback and our preliminary work using AI-assisted data extraction have shown that when AI extracts data, provides a source/rationale for its decision, and skips to the relevant section of a PDF document, it saves time compared to a human extracting data alone. This approach allows the researcher to validate every single data point extracted for accuracy, while saving time from not having to parse as much unnecessary text. We used LLMs to help develop a graphical user interface to facilitate this process, as described below.

### AI-Assisted Data Extraction: A Graphic User Interface for LLM-based Data Extraction

Our first attempt at building software for HIL data extraction was a free, open-source, graphic user interface package for R called AI-assisted Data Extraction (AIDE). AIDE (Figure 4) allows you to use LLMs for extracting data with *mandatory human validation*. AIDE is available for free and open source via GitHub: https://github.com/noah-schroeder/AIDE. The readme contains installation instructions.

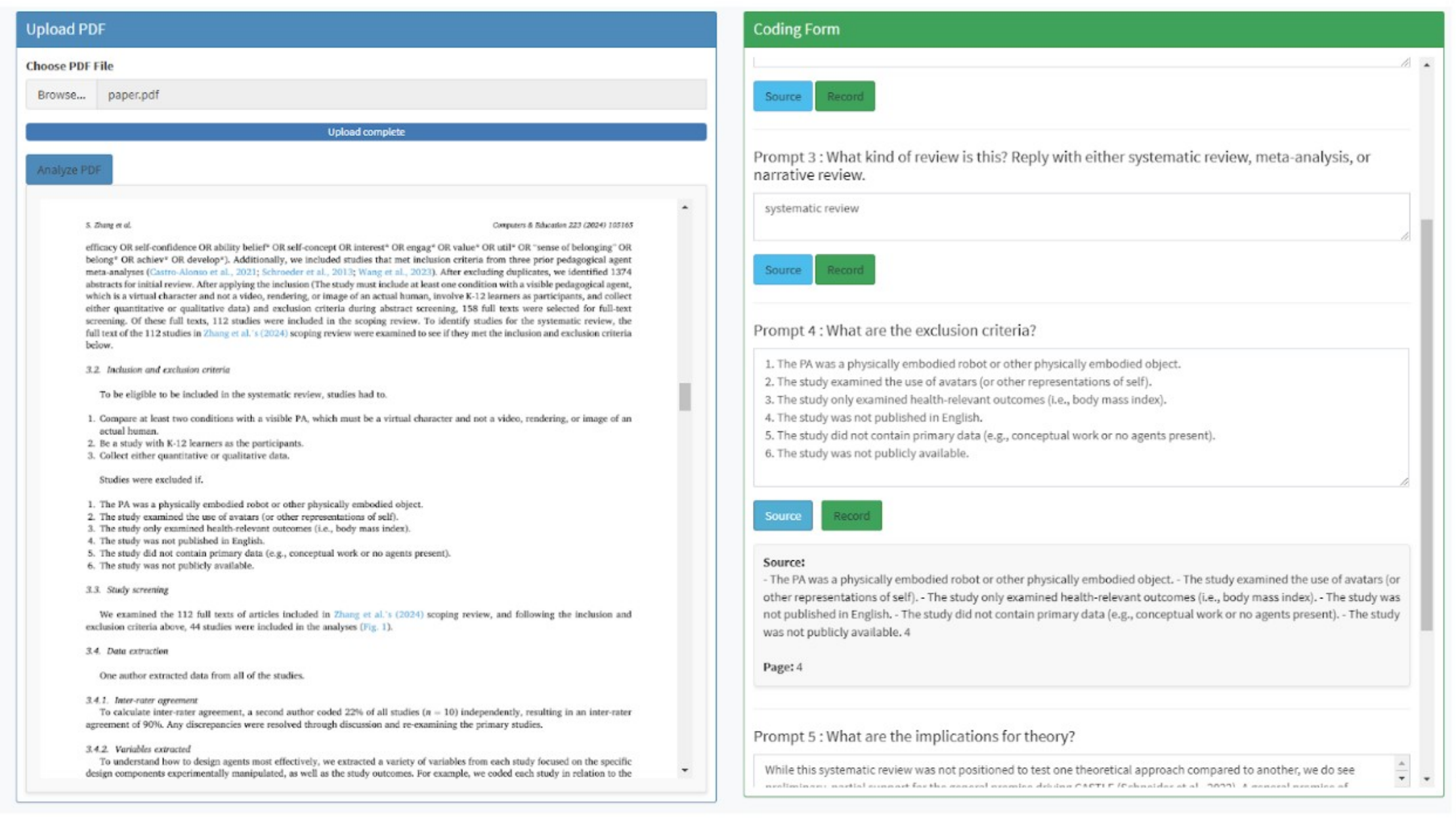

Figure 4. The analysis page within AIDE. Study is presented on the left, while the answers, source text, and record buttons appear on the right.

AIDE was designed to allow you to use free API keys from Google AI Studio, Mistral AI, or Open Router, or use Local models via Ollama. AIDE was designed from the ground up to be free software, so free APIs were prioritized. Note that this pilot version of AIDE was designed to be run locally on your computer, not hosted on an external-facing website (this is noted in the documentation as well). AIDE will launch in a web browser, but by default, it is a local browser window (e.g., localhost) and not public-facing. We do not recommend hosting this R-based version of AIDE on any external-facing website as the API keys would not be secure in that environment.

The user interface for setting up your LLM was relatively straightforward; a user selects their LLM provider, enters their API key, chooses their model, and uploads their coding form. The first row of the coding form will be read as prompts to the LLM and should be written as such. On the analysis page, the user uploads a PDF, which they can see in a PDF viewer. They are automatically presented with an estimate of the number of tokens the API request will use, which allows one to make an informed decision as to if the LLM they plan to use is appropriate. Moreover, this allows one to estimate the cost of their API request if they choose to use a paid LLM. If using local models via Ollama, the user can also select the context window size they want to use.

The right side of the analysis page was auto-populated with the data from the coding form. When "Analyze PDF" is pressed, a single request is made to the LLM API that sends the full text (Google Gemini) or extracted structured full text (Mistral, Open Router, and Ollama), a brief prompt, and the prompts from the coding form. When a response is received, the data is parsed and auto-populated into the corresponding boxes in the user interface. If the Source button is pressed, the LLM's reasoning for the answer is shown when available (see Figure 4), and the PDF reader scrolls to the appropriate page of the PDF automatically. If the answer is correct, pressing the Record button records it to the appropriate location on the coding form on your local computer. If the answer needs to be modified, the user can edit the text before clicking Record. When the user is ready to move to the next paper for analysis, they simply upload a new file, and AIDE will clear the responses in the interface and move to the next row in the coding form file on the local computer. After Analyze PDF is pressed, a request is sent to the API, and the process is identical to the one previously described.

### Limitations of the Pilot Version of AIDE

The pilot version of AIDE is fully functional, but informal stakeholder feedback and our own internal use of the tool revealed three limitations. First, not everyone is familiar with R. This made using AIDE more complicated. Users had to download R, download R Studio, and then use the install scripts provided. This was a notable challenge for non-R users, especially since AIDE was never pushed to CRAN. It became clear that a web-based version of AIDE would be more usable for all end-users.

The second limitation is the hard-coded API endpoints for the various LLM providers. Should endpoints change, or users want to use one that was not listed, the system would no longer function as desired.

The final notable limitation had to do with LLM response parsing. When AIDE was initially developed, we utilized regex to parse the responses so it would split the responses into the appropriate boxes in the user interface (e.g., response, source information). While this worked well for some "families" of LLMs, there were others that would not properly return responses consistent with the required schema. The result was that while some LLMs worked quite well with AIDE, others provided responses that made response parsing difficult.

## AI-Assisted Data Extraction for Web: An Accessible, Open Source Platform for Data Extraction

The initial version of AIDE received anecdotal praise from users, but the limitations noted above ultimately led us to pursue a more easily accessible, web-based version. The web version of AIDE (Figure 5) incorporates all the same key elements to enforce *mandatory human validation* of data extracted. AIDE's web version is freely available on the web: https://noah-schroeder.github.io/AIDE-Web/ and is open source via GitHub: https://github.com/noah-schroeder/AIDE-Web.

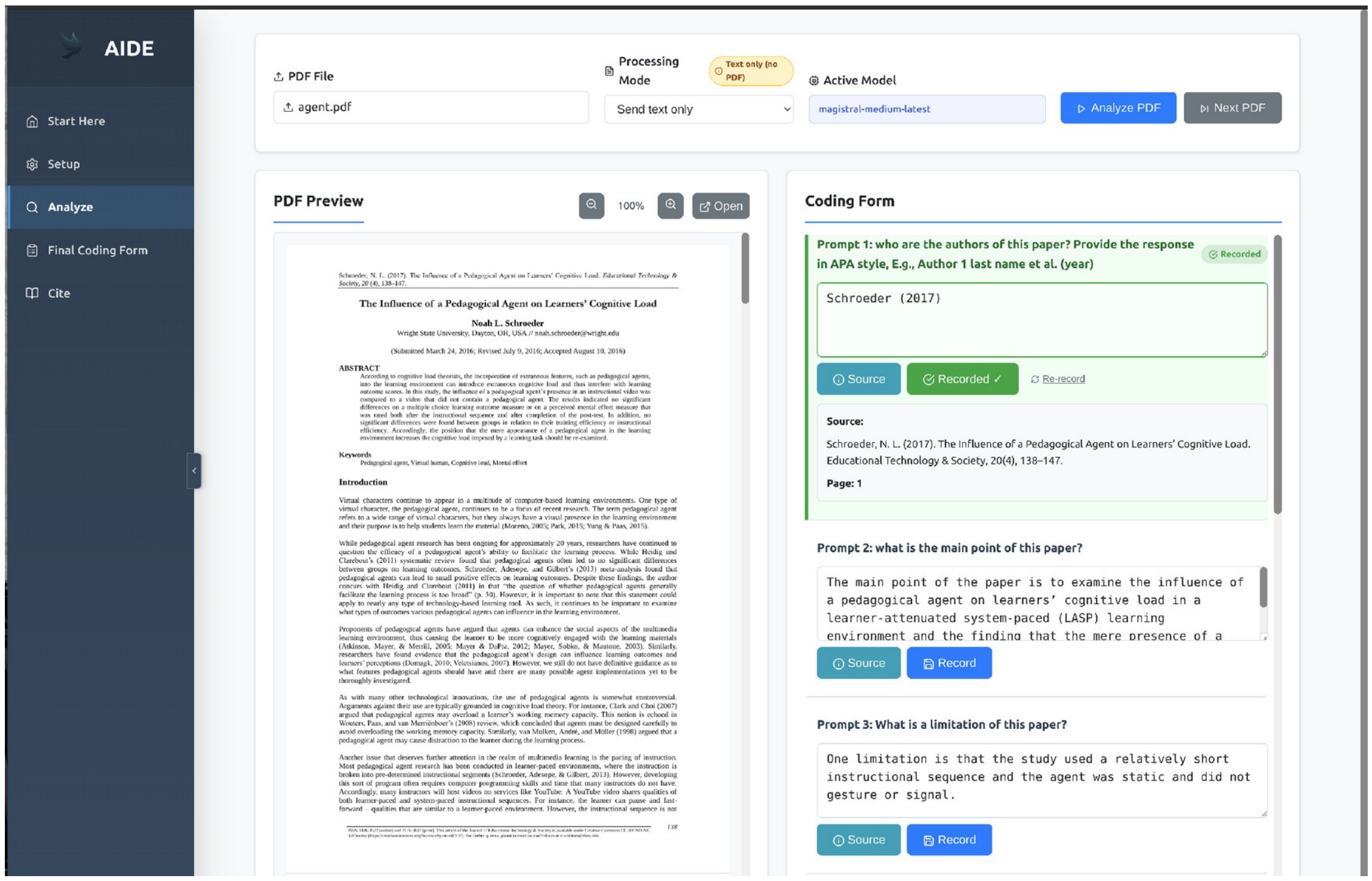

Figure 5. The analysis page within AIDE's web-based version. Note the required human-in-the-loop process to record data and the visual indication when data has been recorded.

### Key Features of AIDE's Web Version

The web version of AIDE incorporates a few key features.

**Security.** API keys are stored in *sessionstorage*, so they are cleared from the local device when the tab is closed.

**LLM Response Parsing.** The web version uses a structured JSON approach to support more consistent response parsing into the appropriate sections of the user interface (e.g., response, source). Moving away from regex to structured outputs has been considerably more consistent within and across various LLM model "families". Specifically, response source information is now much more consistently provided.

**Retaining HIL Validation.** Some users have requested a "record all" function to record all responses at once. We have *not* implemented this because our research (reported above) and that of others (Jansen et al., 2025) shows that LLMs are not yet able to accurately and consistently extract data in a fully automated fashion in a generalizable way. As such, AIDE continues to enforce human validation of every single data point extracted *before* it is ever written into the coding form. We have added a feature to visually show when each data point is recorded to facilitate the user's interaction with the software.

## Implications for HIL Data Extraction in Education

After developing AIDE, we have considered a variety of key questions in relation to its implementation in research synthesis. Below, we address a few common questions researchers may have.

*Should I use HIL data extraction in rigorous research synthesis in education?* We believe it is appropriate to use HIL data extraction in rigorous research synthesis in education if the user is leveraging programs that specifically require every data point to be validated by a human coder *before* it is recorded into a coding form. This prevents reliance on an LLM's decision-making capability (or lack thereof); rather, we are using the LLM to help us extract data faster by focusing our attention on specific parts of the manuscript. If the LLM makes a mistake, it can easily be corrected before recording data to the coding form.

*Does using LLMs for HIL data extraction require explicit acknowledgment and reporting?* Yes, you should explicitly report that generative artificial intelligence was used in the work. Many journals and professional organizations have strict reporting requirements for reporting when generative artificial intelligence is used. Abiding by these requirements is often quite simple, and the explicit nature of the declarations should make it clear to readers that even though LLMs were used to help identify data, in the case of AIDE, humans validated every data point extracted for analysis.

*How should I report that I used HIL data extraction?* At a minimum, we recommend stating the following points in manuscripts where HIL data extraction was used: What software was used for HIL data extraction, what LLM was used, and to what extent humans validated the data extracted.

*Is there anything researchers should know about HIL data extraction?* We encourage researchers to familiarize themselves with the privacy policies of the LLM service providers they choose to use, and note that the privacy policies of the free tier may be different than the privacy policies of the paid tiers.

In addition, AIDE and similar software use what are known as API keys to access web-based LLM services. In practical terms, an API key can be thought of as similar to a password associated with your account to access a service remotely. You should keep these API keys private so that your account is not abused. Many providers will allow you to deactivate or delete an API key when you are finished using it to help prevent future abuses of your account with that key.

*Can open-weight LLMs be used for HIL data extraction?* Absolutely, and we encourage the use of such models as it is easy to document what model was used, at what quantization, and at what parameter settings.

*What about hallucinations, 'context rot', or LLMs referring to information from their 'prior knowledge' rather than the paper in question?* The mere possibility of inaccuracies in data extraction is the reason driving the required human validation that AIDE enforces. When using AIDE, a human must approve every single data point that moves to the coding form, so if irrelevant or inaccurate information is suggested by the LLM, the human can quickly and easily correct the information before it is recorded in the coding form.

*Will we ever be able to fully automate research synthesis in education?* We are unable to predict the future, but we are hesitant to relinquish human oversight of data extraction in a task as consequential as research synthesis. We are reminded of the familiar saying, "Just because you can doesn't mean you should," and find it appropriate to the use of LLM-based data extraction. There are reasons outside of only those discussed here that become relevant if one desired to fully automate data extraction or other research synthesis pipelines, such as bias from LLM training guiding model outputs, but we refrain from discussing them here as they are worthy of more in-depth, dedicated discussions when and if LLMs reach a point where implicitly trusting automation is worth considering for this step of the systematic review process.

## Directions for Future Research

This work highlights a number of potential areas for future research. First, research is needed to quantify the resources saved by using software like AIDE. For example, pertinent outcomes include time saved, workload reduced, and a variety of user perception-related outcomes (e.g., usability, perceived value). This work will help quantify what our team has anecdotally experienced ourselves and heard from colleagues - software like AIDE can save time and effort. However, we do not currently have empirical support for this, only anecdotes.

Second, we encourage researchers to move beyond testing "generalizable" prompts for LLMs, as research shows that they are unlikely to generalize between models (Cao, Sang, et al., 2025) or contexts (Schmidt et al., 2024). Similarly, we discourage researchers from testing models against one another for data extraction tasks and claiming broadly generalizable conclusions because LLMs are sensitive to prompts (Anagnostidis & Bulian, 2024; Errica et al., 2024), context (Schmidt et al., 2024), and parameter settings (Loya et al., 2023), all of which can influence the results. This is particularly the case with closed-source models where opaque changes to underlying models can occur at any time. More fruitful directions may include examining time savings from HIL approaches, fine-tuning and publishing open-weight and open-source models for data extraction tasks in specific contexts, and otherwise investigating how HIL approaches to data extraction can facilitate the research synthesis process.

Finally, there is the question of: what about modern, state-of-the-art frontier LLMs? This will likely be an ongoing question for researchers for the near-term future due to the furious pace at which LLM creators are iterating and launching new models. A consequence of this will likely be the ongoing question, "But can [this new model] do [any specific task] well?" This is, of course, why benchmarks exist. To the best of our knowledge, at this time there are no existing benchmarks for data extraction tasks in systematic review contexts in education. Development of rigorous benchmarks would be an important contribution to the literature to facilitate educational researchers in selecting LLMs to assist their data extraction. One should remember however that benchmarks are simply indicators and LLM creators could train their models on the test items (if the test items are public) to ensure strong performance. For these reasons, our team's perspective, which we feel is consistent with others who have conducted similar methodological work in the field (Jansen et al., 2025), is that in the research synthesis context, human validation of LLM-extracted data remains a mandatory step at this time.

## Conclusion

LLMs offer potential for conserving resources during the data extraction process in research synthesis. Our results showed promise for LLMs being able to accurately extract various types of data, potentially significantly reducing labor time and enhancing the efficiency of research synthesis. However, it became clear that a HIL process (Zanzotto, 2019) is needed to ensure accuracy while still conserving time. As a result, we developed AIDE, a free, open-source program that helps researchers use LLMs to extract and record data in a time-efficient manner.

Research synthesis remains a cornerstone of education, being critically important for policy direction and instructional practice. We must keep the integrity of research synthesis processes as the priority; accuracy should not be sacrificed in the pursuit of cost-savings. While some may question if the "newest" and "best" LLMs perform better than the models we tested here, we in return ask - does it matter? We argue it does not because research synthesis is important enough that human validation of every data point extracted remains a priority.